\documentclass[prl,aps,twocolumn,floatfix]{revtex4-1}
\usepackage{graphicx}
\usepackage{amsmath,amssymb}
\usepackage{dsfont}
\usepackage{xcolor}

\newcommand{\heff}{H_{\scriptstyle \rm eff}}

\newcommand{\hsl}{\mathcal{H}_{\scriptstyle \rm SL}}

\newcommand{\hdd}{\mathcal{H}_{\scriptstyle \rm dd}}

\newcommand{\nn}{\nonumber}

\newcommand{\trl}{\text{Tr}_{ \scriptstyle \rm L}}
\newcommand{\rh}{\rho_{\scriptstyle  S}}
\newcommand{\rl}{\rho_{\scriptstyle \rm L}^{\scriptstyle \rm eq}}

\newcommand{\etal}{\textit{et al.\ }}

\newcommand{\stoo}{\vert_{t\rightarrow 0}}

\newcommand{\comment}[1]{ }
\begin{document}
\title{Emergence and stability of discrete time-crystalline phases in open quantum systems}
\author{Saptarshi Saha}
\email{ss17rs021@iiserkol.ac.in}
\author{Rangeet Bhattacharyya}
\email{rangeet@iiserkol.ac.in}
\affiliation{Department of Physical Sciences, Indian Institute of Science Education and Research Kolkata,
Mohanpur -- 741246, West Bengal, India}

\begin{abstract}

Here we provide a theoretical framework to analyze discrete time-crystalline phases (DTC) in open quantum
many-body systems. As a
particular realization, we choose a quantum many-body system that exhibits cascaded prethermalization . The analysis uses a fluctuation-regulated quantum master equation.  The master
equation captures the dissipative effects of the drive and dipolar coupling on the dynamics regularized by
the thermal fluctuations.  We find that the dissipators from the drive and the dipolar interactions lend
stability to the dynamics and are directly responsible for the robustness. Specifically, we find that longer
fluctuation correlation time enhances the stability of DTC. Our results are in good agreement with the
experiments. Finally, we show and quantify how the DTC performance degrades with temperature.

\end{abstract}

\maketitle

In the recent past, several groups demonstrated, theoretically as well as experimentally, that driven
quantum many-body systems exhibit a variety of exotic out-of-equilibrium phases \cite{Polkovnikov2011,
eisert_quantum_2015,abanin2017, ueda_2020,yao_discrete_2017}, a prime example being the
discrete time-crystalline (DTC) phase \cite{sacha_time_2018,khemani2019}. Wilczek originally proposed that a quantum system may exhibit a time-periodic echo-like behavior through breaking of the continuous time-translation symmetry; however, the idea was later shown to
be untenable \cite{bruno_impossibility_2013, watanabe_absence_2015}. Later Khemani, Else and
others proposed the concept of the DTC phase, which is characterized by breaking the discrete
time-translation symmetry in the presence of a time-periodic Hamiltonian \cite{Khemani2016,else_floquet_2016}. In this
phase, the periodicity of any system observables is an integer multiple of the period of the drive
Hamiltonian. Most commonly, DTCs exhibit a period-doubling or a sub-harmonic response
\cite{else_floquet_2016}.

More recently, several groups experimentally demonstrated the existence of the DTC phase using Nuclear Magnetic 
Resonance (NMR), trapped ions, and circuit QED
systems \cite{zhang_observation_2017, choi_observation_2017, rovny_observation_2018,
pal_temporal_2018, kyprianidis_observation_2021, beatrez_critical_2023, kesler_observation_2021, gong_discrete_2018, 
taheri_all-optical_2022}.  One common feature for most of the
experiments mentioned above is the application of external drives that contain two non-commuting
Hamiltonians are applied in successive time steps. For examples, 
the experimental demonstrations by Choi \etal and Beatrez \etal use two-pulse schemes on driven dissipative dipolar systems
\cite{choi_observation_2017, beatrez_critical_2023}. 
We note that some groups also reported that the sub-harmonic
response are stabilized by dissipation in DTC \cite{kesler_observation_2021, gong_discrete_2018, taheri_all-optical_2022}.

Two-pulse scheme consists of a spin-locking sequence along the `$x$' direction (the first drive) followed by a rotation along the `$y$'
axis (the second drive) \cite{choi_observation_2017, beatrez_critical_2023}.
Choi \etal used an ensemble of dipolar
coupled nitrogen-vacancy (NV) centers, whereas  Beatrez \etal used a dipolar coupled $^{13}$C nuclear spins in a diamond to demonstrate such a novel sub-harmonic response. Both the cases, it was theoretically observed that a critically slow thermalization or prethermalization occurred in the system  which plays a significant role in stabilizing the DTC phase under
perturbations \cite{ho_critical_2017, beatrez_floquet_2021}. 

In general, for an isolated many-body system, the prethermalization in presence of periodic drive is
theoretically analyzed by Floquet theory (i.e., Floquet prethermalization) \cite{beatrez_floquet_2021,
kuwahara_floquetmagnus_2016, bukov_prethermal_2015, rubio-abadal_floquet_2020, yin_prethermal_2021,
holthaus_floquet_2016}. The effect of the Floquet heating in each cycle can be captured by the second-order
contribution of the periodic drive \cite{rubio-abadal_floquet_2020}. In the presence of local interaction
and high-frequency drive, such heating becomes exponentially slow, which results in a long-lived prethermal
plateau in the system \cite{Abanin2015}.  

 Although such systems are not perfectly isolated systems, they are very weakly coupled with the external
environment with a very-long relaxation time \cite{beatrez_floquet_2021,peng_floquet_2021}. Recently we have provided a dynamical approach to describe the prethermalization in periodically
driven dissipative dipolar systems \cite{saha_cascaded_2023,chakrabarti_2022_creation}. We have used a
recently proposed fluctuation-regulated quantum master equation (FRQME) \cite{chakrabarti2018b}, which
successfully explained the interplay between the secular part of the on-resonant periodic drive and dipolar
interaction that led to a prethermal phase. The dynamics is constrained by a set of quasi-conserved
quantities in this regime. Subsequently, the non-secular interactions and system-bath interaction provide a
very slow thermalization process. Thus, the second-order terms of the drive and dipolar interaction
regulated by thermal fluctuations conveniently explain the effect of the Floquet heating in the system. 

The analytical form of the FRQME is given as \cite{chakrabarti2018b},
\begin{eqnarray} \label{frqme}
&&\frac{d\rh}{dt}= -i \trl\Big[\heff(t),\rh \otimes\rl\Big]^{\rm sec}\nn\\
&&-\int\limits^{\infty}_0 d\tau \trl\Big[\heff(t),\Big[\heff(t-\tau),\rh \otimes\rl\Big]\Big]^{\rm
sec}e^{-\frac{\tau}{\tau_c}},
\end{eqnarray} 
where, $\rh$ is the reduced density matrix of the system, $\rl$ is the equilibrium density matrix of the
bath, `$\rm sec$' denotes secular approximation, $\tau_c$ is the fluctuation correlation time-scale, and 
$\heff$ contains all Hamiltonians including drives, dipolar and system-environment couplings. The 
unique feature of the above Eq. (\ref{frqme}) is the presence of the exponential kernel in the second-order
terms that help calculate the dissipative effects of the coupling and drive. FRQME
has been used as an efficient tool for analyzing several applications in NMR, quantum optics, and quantum
information processing \cite{chanda2020,chatterjee_nonlinearity_2020,chanda2021,saha2022,saha_effects_2022}.

Here, we consider a dipolar coupled two-spin-1/2 open quantum system, as a simple realization of the dilute samples used 
in the recent experiments demonstrating DTC using NMR \cite{choi_observation_2017,beatrez_critical_2023}. The secular part of the dipolar
coupling Hamiltonian is denoted by $H_{\scriptstyle \rm dd} = \omega_{d} 
\left(2 I_z^1 I_z^2 - I_x^1 I_x^2 - I_y^1 I_y^2 \right) $. Here, $\omega_{d}$ is the dipolar coupling 
strength. and $I_{\alpha}^i = \sigma_{\alpha}^i/2, \,\alpha \in \{x,y,z\}$, $\sigma_{\alpha}$ is $\alpha$th component of Pauli 
spin operators for $ith$ spin-1/2 particles.
\begin{figure}[htb]
\includegraphics[width=0.64\linewidth]{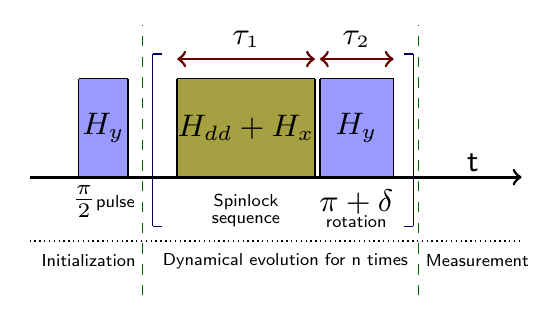} 
\caption{The schematic diagram shows the experimental realization for demonstrating the DTC phase in the
dissipative dipolar system. The initial state is prepared by using a  $\pi/2$ pulse in the  `$y$' direction.
The two-drive protocol is given here. The spin-locking sequence is provided for $\tau_1$ time, secondly, a
$\pi + \delta $ rotation is given along `$y$' direction for a duration $\tau_2$. The whole sequence, $\tau =
\tau_1 + \tau_2$ is repeated for $n$ times before measuring the final spectrum of $S(M_x(t))$.}
\label{fig-1}
\end{figure}

Initially, the spins are oriented along the `$z$' direction in the presence of the
Zeeman field ($\omega_\circ$ is the Zeeman frequency). A $\pi/2$ pulse along the `$y$' direction rotate along the  `$x$'
direction. The initial density matrix is written as, $\rh \stoo = \vert\psi\rangle \langle \psi \vert $. Here, $\vert\psi\rangle
\stoo = \otimes_{i} \left(\vert\uparrow \rangle_i + \vert\downarrow \rangle_i\right)/ \sqrt{2}$.
$\vert\uparrow \rangle$ and $\vert\downarrow \rangle$ are the eigenstate of the Zeeman basis. Next, a
resonant periodic drive along the `$x$' direction is applied for a time duration of $\tau_1$.
The corresponding Hamiltonian in the interaction frame is given by, $H_x = \sum\limits_{i=1}^2 \omega_1 I_{x}^i$. As we are interested in the non-equilibrium dynamics of the system in the presence of the
periodic drive a weak system-bath coupling is chosen. For this system, the short-term  temporal dynamics is dominated by
decay from other sources, i.e., dipolar dissipation and drive-induced dissipation.

Using FRQME (Eq. (\ref{frqme})), the dynamical equation for this case is written as \cite{saha_cascaded_2023},
\begin{eqnarray}\label{dynamics-1}
\frac{d\rh}{dt}&=& \left( \mathcal{D}_{\scriptstyle \rm sec} + \mathcal{D}_{\scriptstyle \rm nsec} 
+ \mathcal{D}_{\scriptstyle \rm SL}\right) [\rh(t)],\quad [0<t<\tau_1]
\end{eqnarray}
where,
$\mathcal{D}_{\scriptstyle \rm sec},\,\mathcal{D}_{\scriptstyle \rm nsec}, \, \mathcal{D}_{\scriptstyle \rm SL} $ are the contribution coming from the secular terms of drive and dipolar interaction, the non-secular terms, and the system-bath coupling respectively.
We choose, $\omega_1, \omega_d \gg \omega_{\rm SL}$ ($\omega_{\rm SL}$ is system-bath coupling strength) and $\omega_\circ \tau_c >1$. Hence, $\mathcal{D}_{\scriptstyle \rm sec}[\rh]>\mathcal{D}_{\scriptstyle \rm nsec}[\rh] > \mathcal{D}_{\scriptstyle \rm SL}[\rh]$. Therefore the contributions of the last two terms are ignored. The analytical form of $\mathcal{D}_{\scriptstyle \rm sec}[\rh]$ is given as,
\begin{eqnarray}
\mathcal{D}_{\scriptstyle \rm sec}[\rh] = -i\left[H_{\scriptstyle \rm sec}, \rh\right] -  \tau_c\left[
H_{\scriptstyle \rm sec},\left[H_{\scriptstyle \rm sec},\rh \right] \right],
\end{eqnarray}
 Here, $H_{\scriptstyle \rm sec}$ is the interaction representation of the secular part of $H_x + H_{\scriptstyle \rm dd} $. We analyze Eq. (\ref{dynamics-1}) using a set of symmetric and anti-symmetric observables, for which we have.
\begin{eqnarray}\label{obs}
\rh (t) &=& \sum\limits_{\alpha, \beta }  A_{\alpha \beta} (t)I_{\alpha} \otimes I_{\beta},
\end{eqnarray}
where, $\alpha$, $\beta \in \{x,y,z,d\}$, and $I_d = 2\times 2$ identity matrix.  $M_i = A_{id}  + A_{di}$, $M_{ii} = A_{ii}$, $M_{ij} = A_{ij} + A_{ji}$ $\forall i\neq j$ and $i,j \neq d$. In terms of observables, the dynamics can be divided into two subgroups $\{M_x,\, M_{yy}, M_{zz}, M_{yz}\}$ and  $\{M_z,\, M_{y}, M_{xz}, M_{xy}\}$. For the given initial condition the dynamics is confined only within the first subgroup. Several conserved quantities constrain the dynamics which can be written using the linear combination of the observables from the first subgroup. Their forms are given as $3\omega_{d}\dot{M}_{zz}+\omega_1  \dot{M}_{x} = 0$,  
$\dot{M}_{yy}+\dot{M}_{zz} =0$ and $\dot{M}_{xx} = 0$ \cite{saha_cascaded_2023}. The solution of the above
Eq. (\ref{dynamics-1}), in terms of the observable, is denoted as $M_i^{\scriptstyle \rm pre}(\alpha, t)$. Here `pre' denotes the prethermal state, and $\alpha$ is the initial value of the observable.  The value of $M_x(t)$ in the prethermal state is written as, 
\begin{eqnarray}
M_x^{\scriptstyle \rm pre}(M_\circ,t_{\scriptstyle \rm pre})  = M_\circ \frac{\omega_1^2}{\omega_1^2 + 9\omega_{d}^2/16}
\label{mxpre}
\end{eqnarray}
Here, $M_x \stoo = M_\circ$ and $t_{\scriptstyle \rm pre}$ denotes the time required to reach the prethermal state.
In experiments, drive strength is taken to be stronger than the dipolar interaction ($\omega_1 > \omega_{d}$). Such consideration leads to the negligible decay of $M_x(t)$ in the transient phase \cite{choi_observation_2017}. 

After the spin-locking sequence, a $\pi+ \delta$ rotation is applied about the $y$ direction. The required time $\tau_2$ is usually small and hence the drive-induced dissipation are ignored \cite{beatrez_critical_2023}. 
Here, $H_y = \omega_2 \sum_{i=1}^2 I_y^i$ and $\omega_2\tau_2 = \pi + \delta$. To obtain the $2\tau$ periodic response in the system, the total sequence, $\tau = \tau_1 + \tau_2$ is repeated for $n$ times $[n \in \mathcal{I}]$. The final density matrix $\hat{\rh}(t)$ after the $n$-cycle can be written as,
\begin{eqnarray}\label{main-eq}
\hat{\rh}(t) =\left[e^{\hat{\mathcal{L}}_y  \tau_2} e^{\hat{\mathcal{L}}_{\scriptstyle \rm sec}  \tau_1}\right]^n \hat{\rh} \stoo.
\end{eqnarray}
Here, $\hat{\mathcal{L}}_y$ is the Liouvillian corresponding to $H_y$. We also define, $\hat{\mathcal{L}}_1$
is the Liouvillian corresponding to $H_{\scriptstyle \rm dd} + H_x$, therefore,
$\hat{\mathcal{L}}_{\scriptstyle \rm sec} = \hat{\mathcal{L}}_1 + \tau_c \hat{\mathcal{L}}_1 \times\hat{\mathcal{L}}_1 $.

\begin{figure}[htb]
\includegraphics[width=\columnwidth]{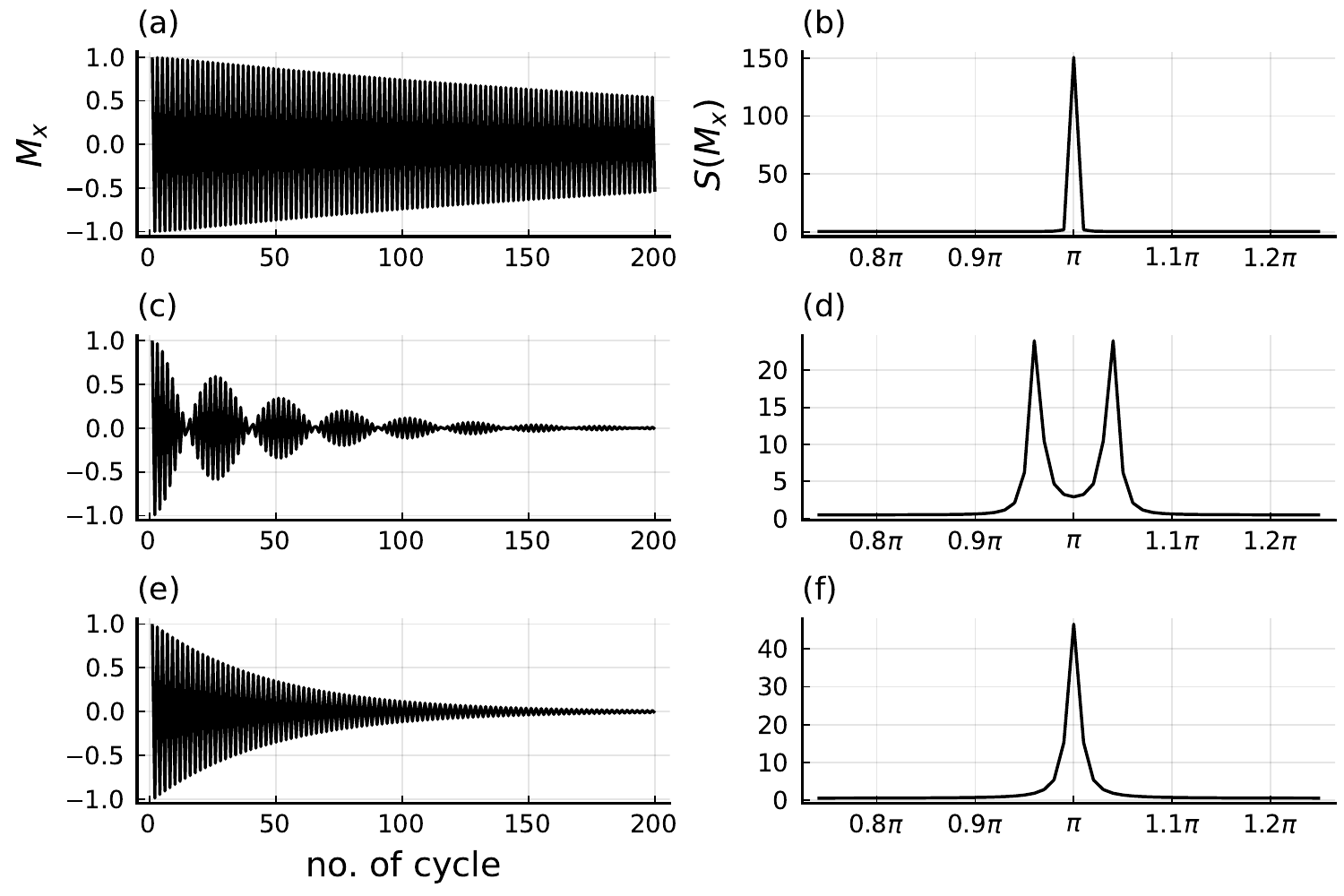} 
\caption{Plot of $M_x$ versus the number of cycles is shown in  (a), (c),  (e) and their corresponding
Fourier transform, $S(M_x)$ versus $\omega$ are shown in  (b), (d), and (f). The value of the fixed
parameters are given as, $\omega_1 = 2\pi \times 50$ KHz, $\omega_2 = 2\pi \times 100$ KHz, $\omega_{d} =
2\pi \times 2$ KHz, $\tau_c = 10^{-3}$ ms, and $n=200$. For the
upper plot (Fig. 2 (a), (b)), the value of $\omega_1 \tau_1 = 2\pi\times 0.02$, $\omega_2 \tau_2 = \pi$.
For the middle plot (Fig. 2 (c), (d)), the value of $\omega_1 \tau_1 = 2\pi\times 0.02$, $\omega_2 \tau_2
= 1.04\pi$. For the lower plot (Fig. 2 (e),(f)), the value of $ \omega_1 \tau_1 = 2\pi$, $\omega_2\tau_2 =
1.04\pi  $. The period-doubling occurs for the upper plot (Fig. 2(b)) as the peak of the Fourier spectra
appears at $\omega = \pi$. A small increase in $\omega_2 \tau_2$ destroys the $2\tau$ periodic response in
(shown in Fig. 2(d)), which can be retrieved for a significant increment of $\omega_1\tau_1$ (shown in Fig.
2(f)). }
\label{fig-2}
\end{figure}

\begin{figure*}[htb]
\includegraphics[width=0.7\linewidth]{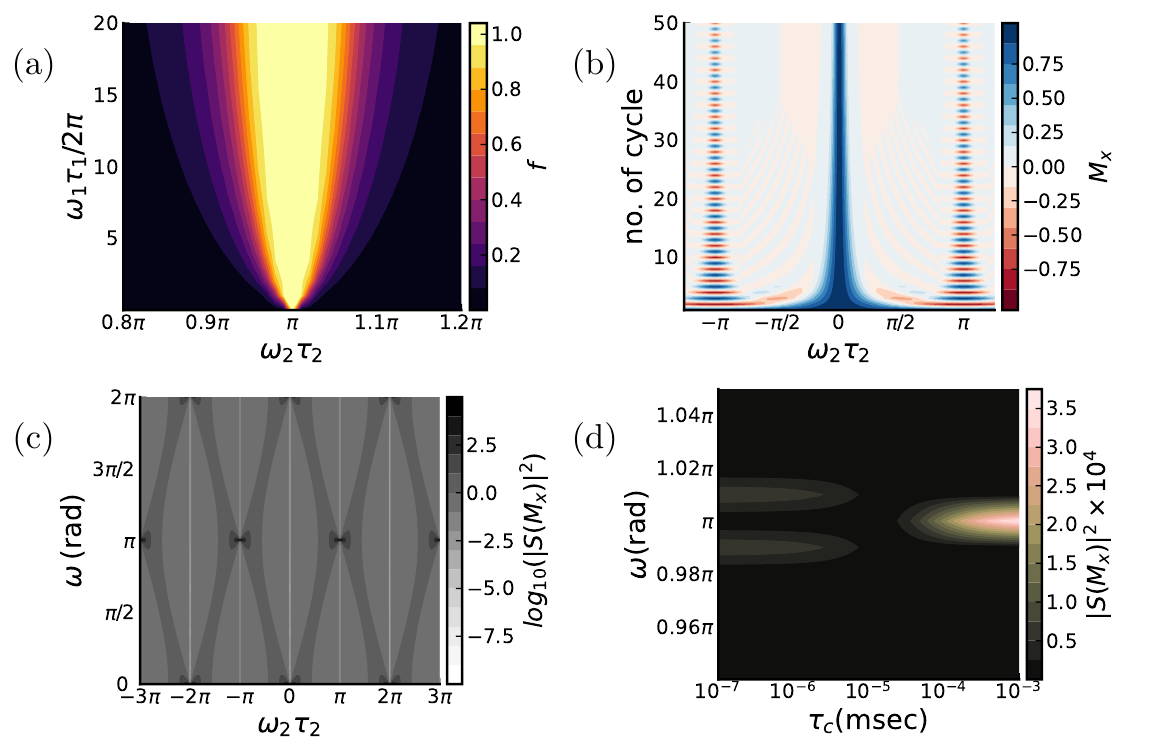} 
\caption{Fig. (a) shows the contour plot of the crystalline fraction $(f)$ as a function of $\omega_1
\tau_1$ and $\omega_2 \tau_2$. The list of fixed parameters are given here, $\omega_1 = 2\pi \times 20$ KHz,
$\omega_2 = 2\pi \times 1000$ KHz, $\omega_{d} = 2\pi \times 0.2$ KHz, $\tau_c = 10^{-4}$ ms, and $n=200$.
We note that the DTC phase is more robust for lower values of $\omega_2 \tau_2$ and higher values of
$\omega_1 \tau_1$. Here $f = 0.1$ is defined as the phase boundary. Fig. (b) shows the contour plot of $M_x$
as a function of $\omega_2 \tau_2$ and the cycle number. The list of fixed parameters are given here,
$\omega_1 = 2\pi \times  20$ KHz, $\omega_2 = 2\pi \times 1000$ KHz, $\omega_{d} = 2\pi \times 2$ KHz,
$\tau_c = 10^{-3}$ ms, $\omega_1 \tau_1 = 2\pi $, and $n=50$. The blue region around $\omega_2
\tau_2 = 0$ is known as the prethermal phase. The alternative blue and red stripes around $\omega_2 \tau_2 =
\pm \pi$ show the existence of the DTC phase. Fig. (c) shows the contour plot of $\vert S(M_x) \vert^2$ as a
function of $\omega_2 \tau_2$ and $\omega$. The list of fixed parameters are given here, $\omega_1 = 2\pi
\times  20$ KHz, $\omega_2 = 2\pi \times 1000$ KHz, $\omega_{d} = 2\pi \times 0.5$ KHz, $\tau_c = 10^{-3}$
ms, $\omega_1 \tau_1 = 2\pi \times 2  $, and $n=200$. The Fourier peak around $\omega = \{0,\,
2\pi\}$ arises due to the presence of the prethermal phase; similarly, the peaks around $\omega = \pi$
denote the emergence of the DTC phase.  Fig. (d) shows the contour plot of $\vert S(M_x) \vert^2$ as a
function of $\tau_c$ and $\omega$. The list of fixed parameters are given here, $\omega_1 = 2\pi \times  50$
KHz, $\omega_2 = 2\pi \times 1000$ KHz, $\omega_{d} = 2\pi \times 0.1$ KHz, $\omega_1 \tau_1 = 2\pi \times 2 
$, $\omega_2 \tau_2 =   1.01 \pi$, and $n=200$. The DTC phase is more
robust in the high $\tau_c$ regime. Lowering $\tau_c$ leads to the breaking of $2\tau$ periodicity in the
system.}
\label{fig-3}
\end{figure*}

We numerically solved Eq. (\ref{main-eq}) and calculate $M_x(t)$ and its spectrum for the initial condition
$M_x\stoo = 1$ and plot in Fig. \ref{fig-2}. For a fixed $\omega_{d},\, \omega_1,\, n$ and $\tau_c$,
when $\omega_2 \tau_2 = \pi$ and $ \omega_1\tau_1 = 2\pi\times 0.02 $, we confirm a sub-harmonic response. For $\omega_2 \tau_2  = \pi$, the `$x$' magnetization flipped from $\hat{x}$ to $-\hat{x}$ in each
cycle, therefore the spectrum has a single peak at a position $\omega = \pi$ (upper curve in Fig.
\ref{fig-1}). Next, we study the stability of the $2\tau$ response by varying the experimental parameters
$\tau_1$ and $\delta$. For the same $\tau_1$, if $\delta = 0.04 \pi$, the $2\tau$ periodicity has vanished,
and we get two peaks very close to $\omega = \pi$. Such period-doubling response can be retrieved with a
larger $\tau_1$ (100 times than the previous one). Therefore, our result suggests that lower $\delta$ and
higher $\tau_1$ are desirable for observing the DTC phase. Following the works by Choi \etal, we define the
crystalline fraction ($f$) as, $f = \vert S(\omega = 0.5)\vert^2 / \sum_\omega \vert S(\omega )\vert^2$
\cite{choi_observation_2017}. We also show a contour plot of crystalline fraction ($f$) as a function of
$\omega_2 \tau_2 $ and $\omega_1 \tau_1$  to capture the regime of the DTC phase in the system. The plot of
$f$ clearly shows the dependence of $\delta$ and $\tau_1$  on the DTC phase. The stable regime is shown in
the yellow color of the contour plot in Fig. \ref{fig-3}(a),  which also matches with earlier experimental results
\cite{choi_observation_2017}. Here $f = 0.1$ is defined as the phase boundary, so, below $f = 0.1$, there is
a smooth crossover from the DTC phase to non-DTC phase.

For $\omega_2\tau_2 = 0$, the solution of the dynamical equation (Eq. (\ref{main-eq})) shows the
prethermalization in the system as $M_x (t)$ is constant (Eq. (\ref{mxpre})). Therefore, in Fig.
\ref{fig-3}(b), prethermalization occurs in the blue regime (color online) near the $\omega_2\tau_2 = 0$. On
the other hand, the system reaches the DTC phase around  $\omega_2\tau_2 = \pm \pi$. The alternative dark bands occur due to the flipping of $M_x$ at each cycle, $[\text{sgn}[(M_x)_{2n +1}] =
-\text{sgn}[(M_x)_{2n}]]$, which shows  a robust sub-harmonic response around $\omega_2\tau_2 = \pm \pi$.

 To confirm the existence of the DTC phase in the system, we also provide a contour plot of the spectrum of
$M_x$ in Fig. \ref{fig-3}(c).  The prethermal phase occurs when $\omega_2\tau_2 = 2N \pi$, $\{N \in I\}$.
Therefore, the Fourier peak arises at $ \omega = 0, \, 2\pi$ for that particular values of $\omega_2
\tau_2$, which signifies that the applied drive and the response have the same periodicity at the prethermal
phase. From Fig. \ref{fig-3}(b), we find that, DTC phase occurs at $\omega_2 \tau_2 = \pm \pi$, similarly
the Fourier peaks in Fig. \ref{fig-3}(c), arises at $\omega =  \pi$ for $\omega_2 \tau_2 \approx \pm (2N +1)
\pi$.  The results corresponding to Fig.
\ref{fig-3}(b), (c) are in agreement with the experimental observation by Beatrez \etal
\citep{beatrez_critical_2023}. 

In our dynamical equation (Eq. (\ref{main-eq})), $\tau_c$ plays an important role. To demonstrate the
$\tau_c$ dependence of the DTC phase, we provide a simplified analytical calculation (a detailed version 
of the calculation provided in the supplementary article). Here, we assume that
$\delta/ \pi \ll 1 $, so, $\sin(\pi +\delta) \approx -\delta, \, \cos(\pi +\delta)  \approx -1$. In table
\ref{table-1}, the solution of the dynamical equation (Eq. (\ref{main-eq})) is provided for each time
instance $\tau_1, \, \tau_2$ up to the period $2\tau$. Here, $\omega_2 \tau_2 = \pi + \delta$.
\begin{table}[htb]
\begin{center}
\caption{Solution of $M_x (t)$ at every time instances.}
\begin{tabular}{ p{1.5cm} |p{4.5cm}    }
 Time$(t)$& $M_x (t)$\\
 \hline
 \hline
 $0$ & $M_{\circ}$\\
 $\tau_1$  &  $M_x^{\scriptstyle \rm pre}(M_\circ,\tau_1)$    \\
 $\tau $  & $-M_x^{\scriptstyle \rm pre}(M_\circ,\tau_1)$     \\
 $\tau + \tau_1$  &  $-M_x^{\scriptstyle \rm pre}(M_x^{\scriptstyle \rm pre}(M_{\circ}, \tau_1), \tau_1)$    \\
 $2\tau$  &  $ M_x^{\scriptstyle \rm pre}(M_x^{\scriptstyle \rm pre}(M_{\circ}, \tau_1), \tau_1) +$    \\
 &   $\delta^2 M_z^{\scriptstyle \rm pre}(M_x^{\scriptstyle \rm pre}(M_{\circ}, \tau_1), \tau_1)$ \\
 \hline
  \end{tabular}
\label{table-1}
\end{center}
\end{table} 
The solution shows that the `$x$' magnetization is flipped in every first cycle of time-period $\tau$. In
the next cycle, $M_x$ doesn't flip perfectly, as $\delta \neq 0$. Therefore, a small imperfection in the
$\pi$ rotation along the `$y$' direction destroys the DTC order in the system. Although a sufficiently long
$\tau_1$ helps to bring back the DTC order because for large $\tau_1$. The extra term of $M_x (2\tau)$ in
the table \ref{table-1} decays to zero as in the limits of $\omega_1> \omega_{d}$, the analytical form of
the extra term is written as $M_z^{\scriptstyle \rm pre}(\alpha, \tau_1) \approx \alpha\cos(\omega_1
\tau_1) e^{-\omega_1^2 \tau_1 \tau_c}$. Hence the solution becomes, $M_x (2\tau) = M_x^{\scriptstyle
\rm pre}(M_x^{\scriptstyle \rm pre}(1, \tau_1), \tau_1) $  for higher $\tau_1$. We find that the roles of $\tau_1$ and $\tau_c$ are
complementary. It is well known that $\tau_c$ is inversely proportional to the temperature
\cite{chakrabarti2018a}. Hence, at lower temperature, the effect of the imperfect rotation would be
diminished, and the sub-harmonic response could be restored. We also show a contour plot of $\vert
S(M_x)\vert^2$ as a function of $\tau_c$ and $\omega$, which is equivalent to the analytical expression,
which is provided in table \ref{table-1}. Therefore, the $2\tau$ periodic response vanishes for lowering
$\tau_c$.

In the $\tau_1$ time duration, the dynamics are governed by both first-order unitary and second-order
dissipative processes. Such dissipation is essential for the persistence of DTC order in the system. Due to
the second-order terms, the extra effects due to the imperfect rotation in $\tau_2$ time duration are
reduced. In addition to that, $M_x$ is stabilized by the dissipation,  as the dissipation doesn't destroy
the conserved quantity  $(\omega_1\dot{M}_x + 3 \omega_{d}\dot{M}_{zz}  = 0)$,  which helps to retain the
DTC phase. A standard QME fails to capture this aspect of the dynamics as DID are not included. Thus, only FRQME can
describe the origin of DTC in a dissipative dipolar system. 
In Fig. \ref{fig-3}(a),
we show that the DTC phase is stabilized for higher $\omega_1 \tau_1$. Although, in real experiments, the
DTC regime becomes narrower for longer values of $\omega_1 \tau_1$
\cite{choi_observation_2017,ho_critical_2017}. Such narrowing occurs due to other sources of
slow decay coming from $\hsl$ and non-secular part of $\hdd$, which we have neglected in this case. 

Recently, 
it has been reported that the interplay between
drive and dissipation leads to persistence oscillation, which plays a key role in the emergence of the DTC
phase in driven dissipative systems \cite{gambetta_discrete_2019,lazarides_time_2020,
buca_non-stationary_2019, passarelli_dissipative_2022,sarkar_signatures_2022}. For example, due to the
imperfect rotation in dissipative-Floquet systems, the system starts to evolve in the wrong sector of the
Hilbert space, which is corrected by dissipation, leads to a stable DTC \cite{lazarides_time_2020}.
Similarly, in our case,  due to imperfect rotation along the `$y$' axis, $M_z$ starts to evolve, which can
be stabilized by increasing the dissipation time ($\tau_1$). Therefore, our results support the recent
theoretical arguments on the stability of DTC in dissipative systems \cite{lazarides_time_2020}.

Finally, our results are in excellent agreement with the experimental observations. 
Moreover, we find that the DTC phase depends on the environmental parameter $(\tau_c)$, which in turn suggests 
that the DTC phase is more stable at low temperatures. We also note that FRQME has a region of validity as
too low temperature might break the timescale separation argument, requiring a non-Markovian approach. We
envisage that our approach will be useful in quantum synchronization problem.


\begin{acknowledgments}
The authors thank Arpan Chatterjee for his insightful comments and helpful suggestions. SS gratefully
acknowledges the University Grants Commission (UGC) of Govt. of India for a research fellowship (Student ID:
MAY2018- 528071).
\end{acknowledgments}

\bibliographystyle{apsrev4-1}
\bibliography{references}
\end{document}


\title{\emph{Supplementary article}: \\
Emergence and stability of discrete time-crystalline phases in open quantum systems}
\author{Saptarshi Saha}
\email{ss17rs021@iiserkol.ac.in}
\author{Rangeet Bhattacharyya}
\email{rangeet@iiserkol.ac.in}
\affiliation{Department of Physical Sciences, Indian Institute of Science Education and Research Kolkata,\\
Mohanpur -- 741246, West Bengal, India}

\maketitle
\section{Dynamical Equation } 
In this section, we discuss the detailed dynamics of the two-spin dipolar coupled system in the presence of
a thermal bath. Following the recent experiments, the system is subjected to a two-pulse driving protocol
which consists of two non-commuting Hamiltonians acting in two successive periods
\cite{beatrez_critical_2023, choi_observation_2017}. The system-bath coupling strength is relatively weaker
than the periodic drive and dipolar interaction. Hence, we neglect the effect of system-bath coupling in our
dynamics. A schematic diagram is provided for a better understanding of the dynamics of this system.
\begin{figure}[htb]
\includegraphics[width=0.48\linewidth]{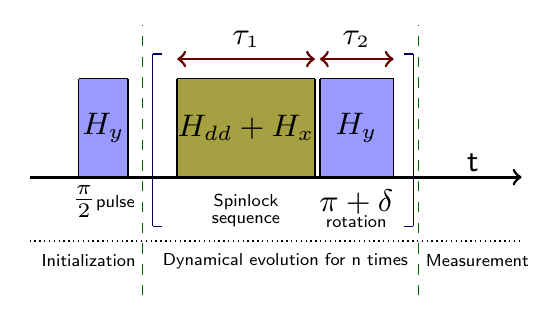} 
\caption{The schematic diagram shows the experiment demonstrating the DTC phase in a dissipative dipolar 
system. The initial state is prepared by using a  $\pi/2$ pulse in the `$y$' direction. Next, the two-pulse protocol begins; it is shown enclosed in the square brackets. The spin-locking pulse is applied for $\tau_1$ 
time, followed by a $\pi + \delta $ rotation pulse applied along `$y$' direction for a duration $\tau_2$. 
The whole sequence, $\tau = \tau_1 + \tau_2$ is repeated for $n$ times before measuring $S(M_x(t))$.}
\label{fig-1}
\end{figure}
\subsection{Dynamical equation during spin-locking pulse} \label{sub1}
The total Hamiltonian of the system is written as,
\begin{eqnarray}\label{ham}
\mathcal{H}(t)&=& \hs^{\circ} + \hl^{\circ}  + \hsl + \hdd + \mathcal{H}_x(t)+\hl(t).
\end{eqnarray} 
The first two terms are the free Hamiltonians of the system.
$\hsl$ represents the weak system-bath coupling. $\hdd$ is the dipolar interaction between the two spins. We 
select only the secular part of the Hamiltonian, assuming that the effect of the nonsecular part is much
smaller. For details, please refer to our earlier work on prethermalization \cite{saha_cascaded_2023}. The analytical 
form of the secular dipolar Hamiltonian is given by
\begin{eqnarray}
\hdd^{\scriptscriptstyle \rm sec} = \omega_{d_0}\left(2 I_z^1 I_z^2 - I_x^1 I_x^2 - I_y^1 I_y^2 \right)
\end{eqnarray}
Here, $\omega_{d_0}$ is the dipolar coupling strength and $I_i = \sigma_i/2, \, \{i \in x,y,z\}$, where $\sigma_i$ is the Pauli spin matrix. The form of the on-resonance periodic drive $\mathcal{H}_x$ is given by,
$\mathcal{H}_x =\sum\limits_{i=1}^2 \omega_1 I_{x}^i  \cos\omega t $. $\hl(t)$ represents the equilibrium thermal 
fluctuations present in the thermal bath. We neglect the contribution from $\hsl$ in our analysis. However,
the fluctuations contribute to drive-induced dissipation and dipolar relaxation.
$\hl(t) = \sum_i f_i(t)\vert \phi_i
\rangle \langle \phi_i \vert$. Here, $\{\vert\phi_i\rangle\}$ are the eigenbases of the environment, and  $f_i$s are the Gaussian stochastic variables with zero mean and standard deviation $\kappa \,
\left(\kappa^2=\frac{1}{\tau_c}\right)$, and $\tau_c$ is defined as the bath correlation time. Fluctuation-regulated 
quantum master equation (FRQME) can be used for such cases to derive the dynamical equation of the system 
\cite{chakrabarti2018b}. In the interaction picture of $ \hs^{\circ} + \hl^{\circ} $, the analytical form of FRQME 
for this case is written as, 
\begin{eqnarray}\label{dynamics-1}
\frac{d\rh}{dt}&=& -i\left[H^{\scriptscriptstyle \rm sec}, \rh(t)\right] -  \tau_c\left[ H^{\scriptscriptstyle \rm sec},\left[H^{\scriptscriptstyle \rm sec},\rh(t) \right] \right],\qquad [0<t<\tau_1]
\end{eqnarray}
Here, $H^{\scriptscriptstyle \rm sec}$ is the interaction representation of the secular part of 
$\mathcal{H}_x(t) + \hdd$. The form of the secular part is given by, $H_x = \sum\limits_{i=1}^2 \omega_1 I_{x}^i$ 
and $H_{\scriptscriptstyle \rm dd} = \omega_{d_0} \left(2 I_z^1 I_z^2 - I_x^1 I_x^2 - I_y^1 I_y^2 \right) $. The 
representation of the symmetric observables is written as
\begin{eqnarray}
M_{\alpha} &=&   {\rm Tr}_s  [ (I_{\alpha}\otimes \mathbb{I} + \mathbb{I} \otimes I_{\alpha})\rho_s ] \nn\\
M_{\alpha\beta}&=&   {\rm Tr}_s [(I_{\alpha} \otimes I_{\beta} + I_{\beta} \otimes I_{\alpha})
\rho_s ],\quad \forall\ \alpha\neq\beta \nn \\
M_{\alpha\alpha} &=&  {\rm Tr}_s [(I_{\alpha}\otimes I_{\alpha})\rho_s]. 
\label{observables}
\end{eqnarray} 
Here, $\alpha, \beta \in \{x,y,z\}$. In terms of those observables, we analyze the above Eq. (\ref{dynamics-1}). Such 
dynamical equations were previously reported by us \cite{saha_cascaded_2023}. In the presence of 
$H^{\scriptscriptstyle \rm sec}$, the dynamical equations can be divided into two subgroups. Each group
contains four observables, $\{M_x, M_{yy}, M_{zz}, M_{yz}\}$ and $\{M_y, M_{z}, M_{xy}, M_{xz}\}$. The
dynamical equations for the first group are given by,
\begin{eqnarray}\label{eq-pre} 
\dot{M}_{x} &=& -\frac{9}{4} \omega_{d_0}^2 \tau_c  M_{x} + 6 \omega_1 \omega_{d_0} \tau_c M_{zz} -6 \omega_1 \omega_{d_0} \tau_c M_{yy} -3\omega_{d_0} M_{yz} \\
\dot{M}_{zz} &=&  \frac{3}{4}\omega_1 \omega_{d_0} \tau_c M_{x} -2\omega_1^2 \tau_c M_{zz} +  2\omega_1^2 \tau_c M_{yy} + \omega_1 M_{yz}  \\
\dot{M}_{yy} &=&  -\frac{3}{4}\omega_1 \omega_{d_0} \tau_c M_{x} +2\omega_1^2 \tau_c M_{zz} -  2\omega_1^2 \tau_c M_{yy} - \omega_1 M_{yz}  \\
\dot{M}_{yz} &=& \frac{3}{4}\omega_{d_0} M_{x} -2 \omega_1 M_{zz} +2\omega_1 M_{yy} -(4 \omega_1^2 + \frac{9}{4}\omega_{d_0}^2 ) \tau_c M_{yz}
\end{eqnarray}
There exist several conserved quantities in this block, which are given by,
\begin{eqnarray}
3\omega_{d_0}\dot{M}_{zz}+\omega_1  \dot{M}_{x} &=& 0\\
\dot{M_{yy}}+\dot{M_{zz}} &=& 0\\
\dot{M}_{xx} &=& 0.
\end{eqnarray}
The existence of such conserved quantities ensures that, if initially, any one of the observables from the first group 
is non-zero, then they have finite values at the steady state. The spin-lock sequence leads to the prethermalization 
in the system \cite{beatrez_floquet_2021,saha_cascaded_2023}. Although, by adding $\hsl$ and other non-secular
contributions, such terms will eventually 
vanish. We are only interested in the dynamics of $M_x$ for a short time compared to the time required for
thermalization (a condition amply satisfied in experiments). The solution 
of $M_x(t)$ from the above Eq. (\ref{eq-pre}) for $M_x \stoo = M_\circ$ is given as,
\begin{eqnarray}\label{mxpre}
M_x^{\scriptscriptstyle \rm pre} (M_\circ, t) = M_\circ \left(\frac{4 \omega_1^2}{\kappa_1^2} + \frac{9}{4}\frac{\omega_{d_0}^2}{\kappa_1^2} \cos (\kappa_1 t) e^{- \kappa_1^2 t \tau_c}\right) 
\end{eqnarray}  
Here, $\kappa_1^2 = 4 \omega_1^2 + \frac{9}{4}\omega_{d_0}^2$.
The above solution (Eq. (\ref{mxpre})) signifies that $M_x(t)$ will reach a steady state, and the steady state value 
is given as $M_x^{\scriptscriptstyle \rm pre}\sts = M_\circ \frac{4 \omega_1^2}{\kappa_1^2}$. This steady
state is a prethermal state \cite{saha_cascaded_2023}. The presence of the conserved quantity also signifies that the 
initial magnetization $(M_\circ)$ is shared between $M_x$ and $M_{zz}$. 
\begin{figure*}[htb]
 \raisebox{3cm}{\normalsize{\textbf{(a)}}}\hspace*{-1mm}
\includegraphics[width=0.36\linewidth]{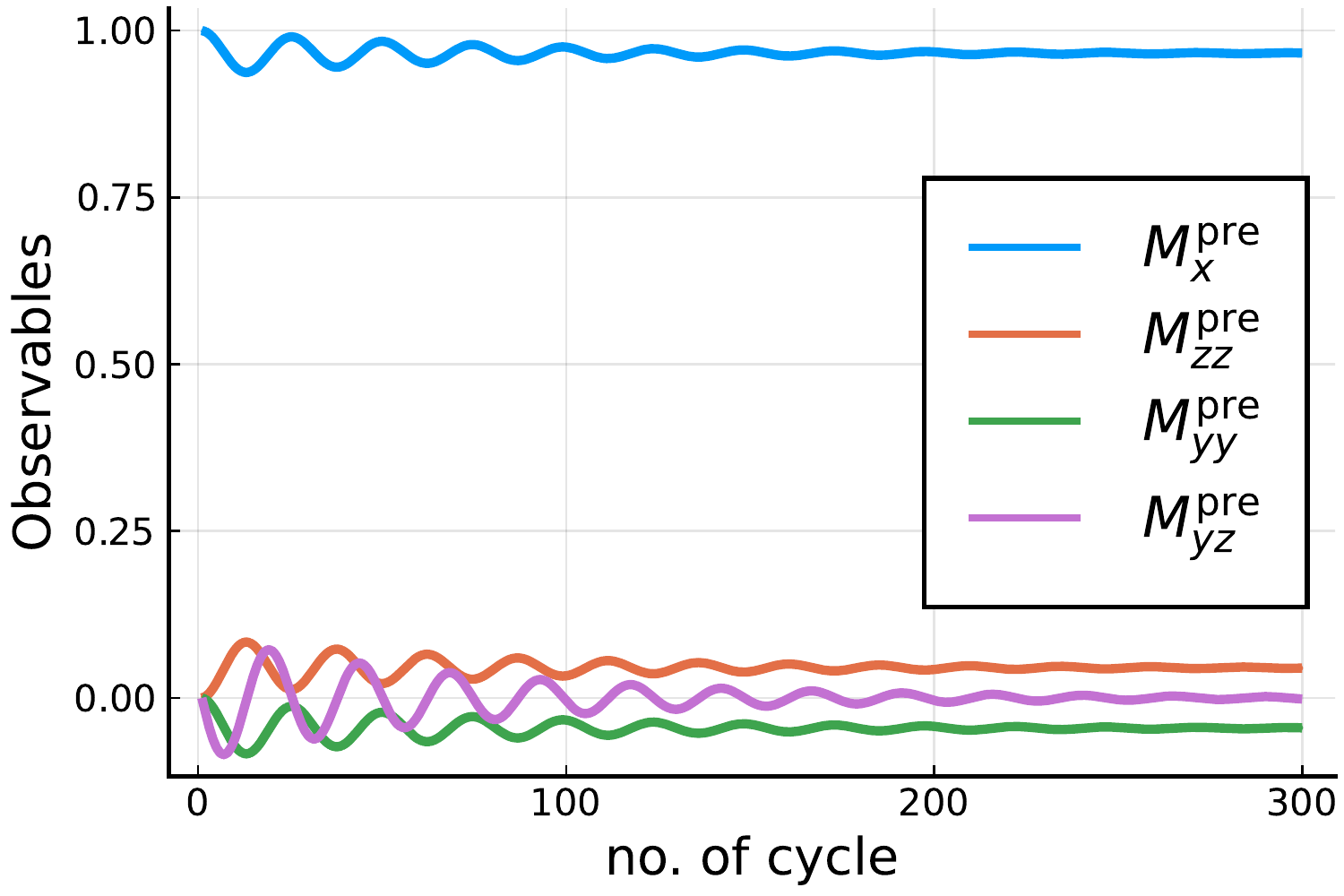} 
\hspace*{3mm}
\raisebox{3cm}{\normalsize{\textbf{(b)}}}\hspace*{-1mm}
\includegraphics[width=0.36\linewidth]{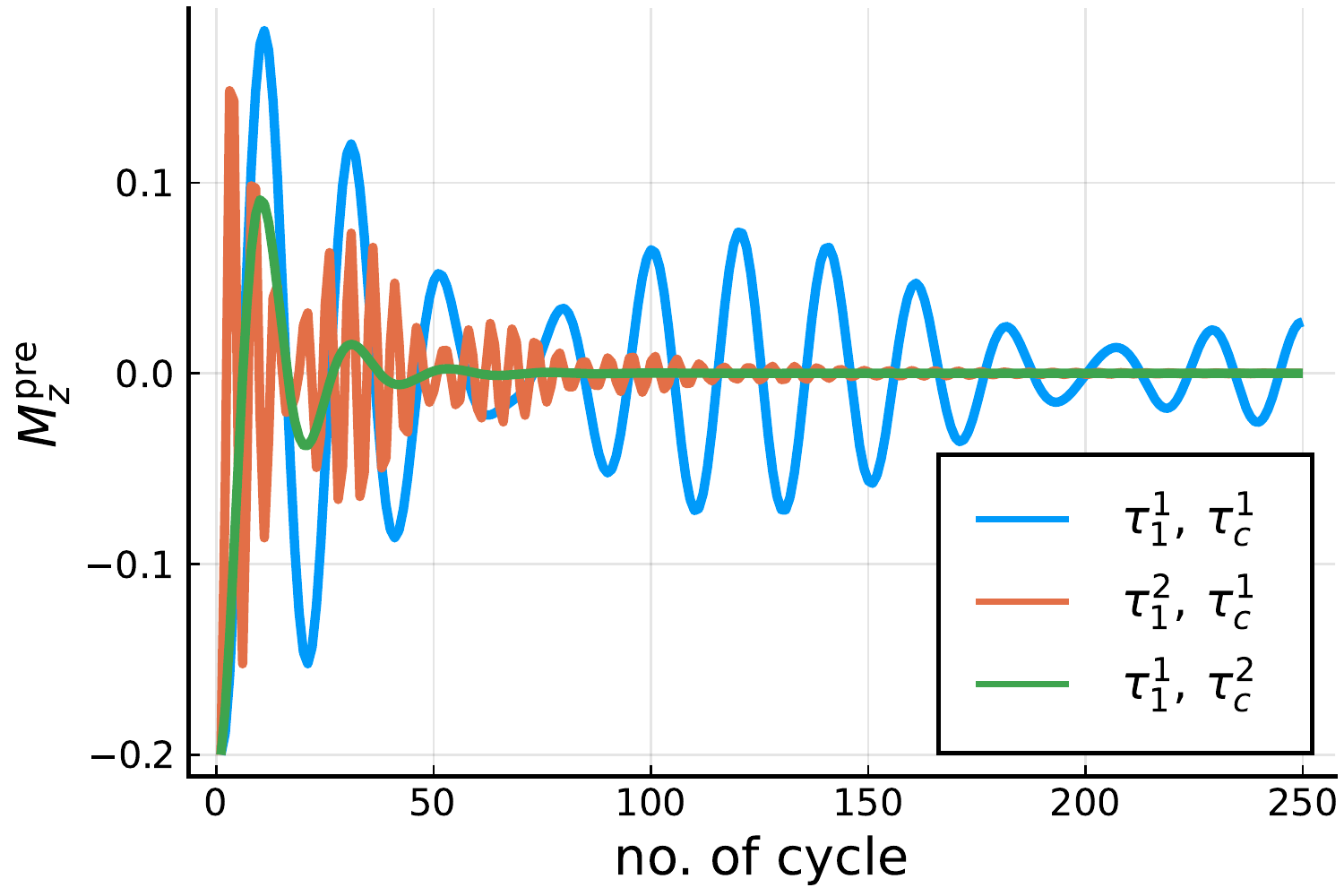} 
\caption{ Fig. (a) shows the plot of observables $M_x^{\scriptscriptstyle \rm pre}, \,
M_{zz}^{\scriptscriptstyle \rm pre}, \, M_{yy}^{\scriptscriptstyle \rm pre}, M_{yz}^{\scriptscriptstyle \rm
pre}$ versus no. of cycle by numerically solving Eq. (\ref{dynamics-1}). Here $\omega_2 \tau_2 = 0$. The
list of fixed parameters are given as, $\omega_1 = 2\pi \times 40$ KHz, $\omega_{d_0} = 2 \pi \times 10$
KHz, $\tau_c = 10^{-4}$ ms and $\tau_1 = 2\pi \times 0.02 / \omega_1$ KHz$^{-1}$. The initial condition is
chosen as $M_x \stoo =1$. For such a choice of parameters, the system reaches a steady state after some
cycle. The dynamical evolution of $M_{zz}^{\scriptscriptstyle \rm pre}, \, M_{yy}^{\scriptscriptstyle \rm
pre}, M_{yz}^{\scriptscriptstyle \rm pre}$ are negligible compared to $M_x^{\scriptscriptstyle \rm pre}$.
Fig. (b) shows the plot of observables $M_z^{\scriptscriptstyle \rm pre}$ versus the number of cycles for three
different choices of $\tau_1, \tau_c$. Here $\omega_2 \tau_2 = 0$. The list of fixed parameters are
$\omega_1 = 2\pi \times 40$ KHz, $\omega_{d_0} = 2 \pi \times 4$ KHz. Three choices are given as $\{\tau^1_1
= 2\pi \times 0.05 / \omega_1$ KHz$^{-1}$, $\tau^1_c = 10^{-4}$ ms $\}$,  $\{\tau^2_1 = 2\pi \times 0.2 /
\omega_1$ KHz$^{-1}$, $\tau^1_c = 10^{-4}$ ms $\}$, $\{\tau^1_1 = 2\pi \times 0.05 / \omega_1$ KHz$^{-1}$, 
$\tau^2_c = 10^{-3}$ ms $\}$. The initial condition is chosen as $M_z\stoo = -0.2$. The above plot 
indicates that for higher values of $\tau_c$ and $\tau_1$, $M_z^{\scriptscriptstyle \rm pre}$ decays faster 
for increasing the number of cycles. Such a condition is necessary for obtaining the DTC phase.}
\label{fig-2}
\end{figure*}
For the demonstration of the DTC phase, the loss of `$x$' magnetization must be very small in each cycle of the 
evolution. Otherwise, $M_x(t)$ will vanish after a few cycles, so we must have $M_x\stoo /M_x\sts \to 1$. Such condition will 
be satisfied if $\omega_1 > \omega_{d_0} $. Choi \etal have chosen the drive and dipolar coupling in such a way that the 
condition is satisfied for their experiment \cite{choi_observation_2017}. From the above plot (Fig. \ref{fig-2}), the 
evolution of $\{M_{zz}, M_{yy}, M_{yz}\}$ are negligible compared to $M_x$ in the limit $\omega_1 > \omega_d$, so for 
the remaining part of the calculation, we ignore their effect on $M_x(t)$. The dynamical equations for the other group 
are written as,
\begin{eqnarray}\label{eq-pre2} 
\dot{M}_{z} &=& - \omega_{1}^2 \tau_c  M_{z} + \omega_1 M_y + 3 \omega_1 \omega_{d_0} \tau_c M_{xz}  \\
\dot{M}_{y} &=& -\omega_1 M_z -(\omega_1^2 + \frac{9}{4}\omega_{d_0}^2 ) \tau_cM_{y} +3 \omega_{d_0} M_{xz} +  3 \omega_1 \omega_{d_0} \tau_c M_{xy}\  \\
\dot{M}_{xz} &=& \frac{3}{4} \omega_1 \omega_{d_0} \tau_c M_{z} -\frac{3}{4} \omega_{d_0} M_{y}-(\omega_1^2 + \frac{9}{4}\omega_{d_0}^2 ) \tau_c M_{xz} + \omega_1 M_{xy}\\
\dot{M}_{xy} &=& \frac{3}{4} \omega_1 \omega_{d_0} \tau_c M_{y} -\omega_1M_{yz} -\omega_1^2 \tau_c M_{xy}
\end{eqnarray}
As there exist no conserved quantities involving the observables from this group, all the observables will vanish at 
the steady state even if they have any initial non-zero values. There is no closed analytical
form for the observables. Hence we provide the plot of the observable $M_z^{\scriptscriptstyle \rm pre}
(M_\alpha, t)$ by numerically solving Eq. (\ref{eq-pre2})  for a different choice of $\tau_1, \, \tau_c$.
Here, $M_z\stoo = M_\alpha$. Other observables are not relevant for this case.
\subsection{Dynamical equation for rotation along the `$y$' direction}\label{sub2}
Next, we consider the rotation along the `$y$' direction. As this is a short-duration pulse, no heating occurs in this 
time duration. In other words, the effects of the dissipators are negligible during this period. Therefore, the dynamics 
is adequately described by the first-order process. The dynamical equation in the interaction frame is given by
\begin{eqnarray}\label{dynamics-2}
\frac{d\rh}{dt}&=& -i\left[H_y, \rh(t)\right],
\end{eqnarray}
Here, $H_y = \omega_2 \sum_{i=1}^2 I_y^i$. 
To obtain the existence of the DTC phase, we are only interested in the dynamics of $M_x$. The solution of Eq. (\ref{dynamics-2}) in terms of $M_x$ is written as,
\begin{eqnarray}
M_x^{\scriptscriptstyle \rm rot} (t) = M_\circ \cos(\omega_2 t),\quad M_z^{\scriptscriptstyle \rm rot} (t) = M_\circ \sin(\omega_2 t) 
\end{eqnarray}
Here $M_x^{\scriptscriptstyle \rm rot} \stoo = M_\circ$. 
\section{Dynamical evolution upto $2 \tau$ time-period}
Using the dynamical equation in the subsections \ref{sub1} and \ref{sub2},  we provide the analytical solution of the 
time evolution of $M_x(t)$ in this section. The dynamical equation in the Liouville space is written as,
\begin{eqnarray}\label{main-eq}
\hat{\rh}(t) =\left[e^{\hat{\mathcal{L}}_y  \tau_2} e^{\hat{\mathcal{L}}_{\scriptscriptstyle \rm sec}  \tau_1}\right]^n \hat{\rh} \stoo.
\end{eqnarray}
Here, $\hat{\mathcal{L}}_y$ is the Liouvillian corresponding to $H_y$. We also define, $\hat{\mathcal{L}}_1$ is the 
Liouvillian corresponding to $H_{\scriptscriptstyle \rm dd} + H_x$, therefore, $\hat{\mathcal{L}}_{\scriptscriptstyle \rm sec} 
= \hat{\mathcal{L}}_1 + \tau_c \hat{\mathcal{L}}_1 \times\hat{\mathcal{L}}_1 $. The time period $(\tau)$ is defined as, 
$\tau = \tau_1 + \tau_2$. For the initial condition $M_x \stoo = 1$, the solution of $M_x(t)$ for 
$\omega_1> \omega_{d_0}$ at time $\tau_1$ is given as, 
\begin{eqnarray}
M_x (\tau_1) = M_x^{\scriptscriptstyle \rm pre}(1, \tau_1),\quad M_z (\tau_1) = 0.
\end{eqnarray}
After the rotation, the solution is given by,
\begin{eqnarray}
M_x (\tau) = M_x^{\scriptscriptstyle \rm pre}(1, \tau_1) \cos\theta, \quad  M_z (\tau) = M_x^{\scriptscriptstyle \rm pre}(1, \tau_1) \sin\theta
\end{eqnarray}
Here, $\theta = \omega_2 \tau_2$. Similarly, after a time $ \tau + \tau_1$, the solution is,
\begin{eqnarray}
M_x (\tau + \tau_1) = M_x^{\scriptscriptstyle \rm pre}(M_x^{\scriptscriptstyle \rm pre}(1, \tau_1), \tau_1) \cos\theta , \quad M_z (\tau + \tau_1) = M_z^{\scriptscriptstyle \rm pre}(M_x^{\scriptscriptstyle \rm pre}(1, \tau_1), \tau_1) \sin \theta
\end{eqnarray}
Finally after the $2\tau$ time-period, the analytical expression of $M_x(2\tau)$ is written as,
\begin{eqnarray}
M_x(2 \tau) = M_x^{\scriptscriptstyle \rm pre}(M_x^{\scriptscriptstyle \rm pre}(1, \tau_1), \tau_1) \cos^2 \theta + M_z^{\scriptscriptstyle \rm pre}(M_x^{\scriptscriptstyle \rm pre}(1, \tau_1), \tau_1) \sin^2 \theta 
\end{eqnarray}
\subsection{Solution for $\theta = \pi$}
Putting $\theta = \pi$, in the above solutions, we get,
\begin{eqnarray}\label{eqpi}
M_x (0) &=&1,\\
M_x (\tau_1) &=& M_x^{\scriptscriptstyle \rm pre}(1, \tau_1),\\
M_x (\tau) &=& -M_x^{\scriptscriptstyle \rm pre}(1, \tau_1),\\
M_x (\tau + \tau_1) &=& -M_x^{\scriptscriptstyle \rm pre}(M_x^{\scriptscriptstyle \rm pre}(1, \tau_1), \tau_1)\\
M_x (2\tau) &=& M_x^{\scriptscriptstyle \rm pre}(M_x^{\scriptscriptstyle \rm pre}(1, \tau_1), \tau_1) 
\end{eqnarray}
In the above Eq. (\ref{eqpi}), we show that, in every time-period $\tau$, the sign of $M_x$ is reversed.
Therefore, $M_x(t)$ comes to the same phase at $2\tau$ time-period. Therefore the system shows the 
period-doubling response, which exhibits the notion of the DTC phase in the system. On the other hand, 
$M_z(t)$ remains zero throughout the evolution.
\begin{figure*}[htb]
\includegraphics[scale =0.9]{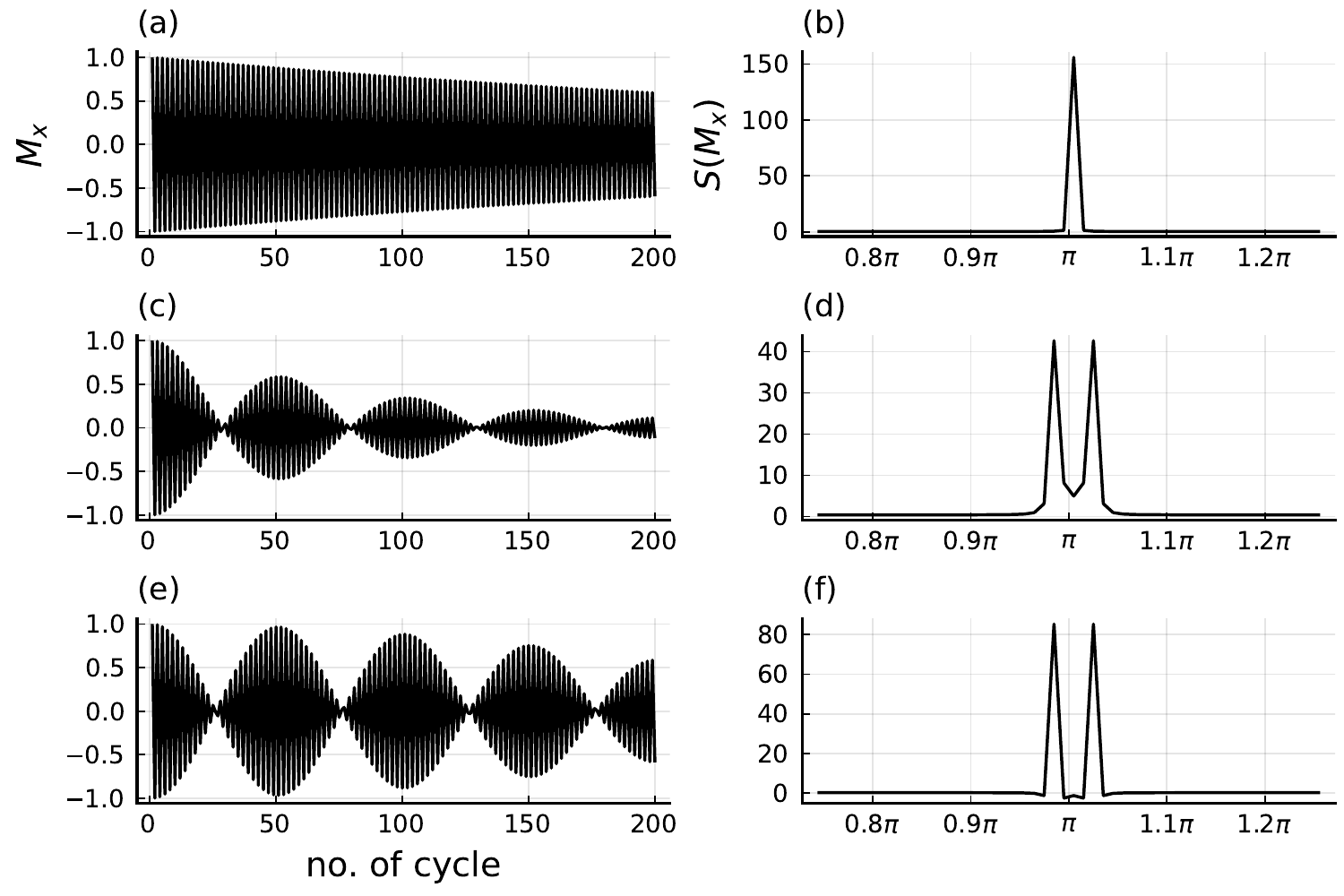} 
\caption{ Plot of $M_x$ versus the number of cycles is shown in  (a), (c),  (e) and their corresponding Fourier 
transforms, $S(M_x)$ versus $\omega$ are shown in  (b), (d), and (f). The chosen parameters are:
$\omega_1 = 2\pi \times 50$ KHz, $\omega_2 = 2\pi \times 100$ KHz, $\omega_{d_0} = 2\pi \times 0.1$ KHz, 
$n=200,$ $\tau_1 = 2\pi\times 0.02/ \omega_1$ KHz$^{-1}$ and $\tau_2 = 1.04\pi/ \omega_2$ KHz$^{-1}$. 
The initial state is chosen as, $\vert\psi\rangle \stoo = \prod_{i=1}^2 \left(\vert\uparrow \rangle_i 
+ \vert\downarrow \rangle_i\right)/ \sqrt{2}$. For the upper plot (Fig. 2 (a), (b); red color online), 
the value of $\tau_c = 10^{-3}$ ms. For the middle plot (Fig. 2 (c), (d); blue color online), the 
value of $\tau_c = 10^{-5}$ ms.  For the lower plot (Fig. 2 (e),(f); green color online), the value 
of $\tau_c = 10^{-7}$ ms. The above plots signifies that lowering $\tau_c$ results in the decaying of 
robust $2 \tau$ response in the system.}
\label{fig-3}
\end{figure*}

\subsection{Solutions for $\theta = \pi + \delta$, with $\delta/ \pi \to 0$}
In such cases, we can assume that, $\cos(\pi + \delta ) = -1$, $\sin(\pi + \delta ) = -\delta$. The solution is given by,
\begin{eqnarray}\label{eqdelta}
M_x (0) &=&1,\\
M_x (\tau_1) &=& M_x^{\scriptscriptstyle \rm pre}(1, \tau_1), \quad M_z (\tau_1) = 0,\\
M_x (\tau) &=& -M_x^{\scriptscriptstyle \rm pre}(1, \tau_1), \quad  M_z (\tau) = -\delta M_x^{\scriptscriptstyle \rm pre}(1, \tau_1)\\
M_x (\tau + \tau_1) &=& -M_x^{\scriptscriptstyle \rm pre}(M_x^{\scriptscriptstyle \rm pre}(1, \tau_1), \tau_1), \quad M_z (\tau + \tau_1) = -\delta M_z^{\scriptscriptstyle \rm pre}(M_x^{\scriptscriptstyle \rm pre}(1, \tau_1), \tau_1)\\
M_x(2 \tau) &=&  M_x^{\scriptscriptstyle \rm pre}(M_x^{\scriptscriptstyle \rm pre}(1, \tau_1), \tau_1)  + \delta^2 M_z^{\scriptscriptstyle \rm pre}(M_x^{\scriptscriptstyle \rm pre}(1, \tau_1), \tau_1)
\end{eqnarray}
In the above solutions (Eq. (\ref{eqdelta})), the $2\tau$ periodicity is absent due to the presence of the 
extra term in $M_x (2 \tau)$. Such a response can be retrieved if $\delta$ is small. From Fig. \ref{fig-2}(b), 
it is clear that, $M_z(M_\alpha,t)$ decays faster for higher values of $\tau_1$. Therefore, higher $\tau_1$ and 
lower $\delta$ are suitable conditions for observing the DTC phase in the system.
\subsection{$\tau_c$ dependency of the DTC phase}
In the limits of $\omega_1 > \omega_d$, the expressions of $M_z^{\scriptscriptstyle \rm pre} (M_\alpha,t)$ is given as,
\begin{eqnarray}
M_z^{\scriptscriptstyle \rm pre} (M_\alpha, t) =  M_\alpha \cos (\omega_1 t)  e^{-\omega_1^2  \tau_c t}
\end{eqnarray}
So, for higher values of $\tau_c$, the effect of the extra term of $M_x (2\tau)$ in Eq. (\ref{eqdelta}) can be 
neglected so, $2\tau$ periodicity is robust in this regime. In the Fig. \ref{fig-2}(b), we show that 
$M_z^{\scriptscriptstyle \rm pre} (M_\alpha, t)$ decays faster for higher values of $\tau_c$, which is
helpful for retrieving the DTC regime. We also plot $M_x (t)$ and it's Fourier transform $\mathcal{S}(M_x (t))$. 
For low values of $\tau_c$, there is no existence of $2\tau$ periodicity, but for changing the value of $\tau_c$ 
from $10^{-7}$ ms to $10^{-3}$ ms, such a novel response is retrieved.

\bibliographystyle{apsrev4-1}
\bibliography{references}